\documentclass[reqno,11pt]{amsart}
\usepackage[utf8]{inputenc}
\usepackage{graphicx}
\usepackage{slashed}
\usepackage{amscd}
\usepackage{amssymb}
\usepackage{esint}
\usepackage{enumitem}
\usepackage{comment}
\usepackage{amsmath}
\usepackage{dsfont}
\usepackage[mathscr]{eucal}
\usepackage{epsfig}
\usepackage{pstricks}
\usepackage{pst-all}
\usepackage{pst-all}

\numberwithin{equation}{section}
\allowdisplaybreaks[1]
\definecolor{labelkey}{gray}{.65}

\title[A Collapse Mechanism Without Heating]{A Collapse Mechanism Without Heating}

\author[F.\ Finster]{Felix Finster}
\address{Fakult\"at f\"ur Mathematik \\ Universit\"at Regensburg \\ D-93040 Regensburg \\ Germany}
\email{finster@ur.de}

\author[C.F.\ Paganini]{Claudio F. Paganini \\ \\ November 2025 / August 2026} 
\address{Fakult\"at f\"ur Mathematik \\ Universit\"at Regensburg \\ D-93040 Regensburg \\ Germany}%
\email{claudio.paganini@ur.de}

\newtheorem{Def}{Definition}[section]
\newtheorem{Thm}[Def]{Theorem}
\newtheorem{Prp}[Def]{Proposition}
\newtheorem{Lemma}[Def]{Lemma}

\newcommand{\Thanks}{\vspace*{.5em} \noindent \thanks}
\newcommand{\beq}{\begin{equation}}
\newcommand{\eeq}{\end{equation}}
\newcommand{\Proof}{\begin{proof}}
	\newcommand{\QED}{\end{proof} \noindent}

\newcommand{\la}{\langle}
\newcommand{\ra}{\rangle}

\newcommand{\bbra}{\mathopen{\ll}}
\newcommand{\kket}{\mathclose{\gg}}
\newcommand{\Sl}{\mbox{$\prec \!\!$ \nolinebreak}}
\newcommand{\Sr}{\mbox{\nolinebreak $\succ$}}

\newcommand{\C}{\mathbb{C}}
\newcommand{\R}{\mathbb{R}}
\newcommand{\1}{\mbox{\rm 1 \hspace{-1.05 em} 1}}

\newcommand{\Pdd}{\mbox{$\partial$ \hspace{-1.2 em} $/$}}

\renewcommand{\Tr}{\text{\rm{Tr}}}
\DeclareMathOperator{\tr}{tr}
\renewcommand{\O}{{\mathscr{O}}}

\newcommand\B{{\mathscr{B}}}

\renewcommand{\H}{\mathscr{H}}

\DeclareMathOperator{\Pexp}{Pexp}

\DeclareMathOperator{\supp}{supp}
\newcommand{\Sig}{\mathscr{S}}

\newcommand{\bitem}{\begin{itemize}[leftmargin=2.5em]}
\newcommand{\eitem}{\end{itemize}}

\newcommand{\sint}{\sum  \!\!\!\!\!\!\! \int }

\DeclareFontFamily{OT1}{rsfso}{}
\DeclareFontShape{OT1}{rsfso}{m}{n}{ <-7> rsfso5 <7-10> rsfso7 <10-> rsfso10}{}
\DeclareMathAlphabet{\mycal}{OT1}{rsfso}{m}{n}

\begin{document}
\maketitle

\begin{abstract}
It is shown that the collapse model derived from the theory of causal fermion systems does not give rise to a heating of the probe, up to higher order corrections in the Planck length. Experimental consequences are worked out. The connections to the continuous spontaneous localization (CSL) model, models involving a second-quantized background and the events-trees-histories (ETH) formulation of quantum theory are discussed.
\end{abstract}

\tableofcontents

\section{Introduction} \label{secintro}
In~\cite{collapse} a new collapse mechanism was proposed based on the theory of causal fermion systems.
In the present paper we analyze this model in more detail and show that, in contrast to all other known
collapse models, our mechanism does not give rise to a heating of the probe\footnote{Throughout the paper, by ``probe'' we mean the
system whose collapse is studied, formed of charged fermionic
quantum particles.}
due to the stochastic
background field. In order to work out the physical consequences of this finding, we give a detailed
discussion of collapse phenomena and compare the different models.
Moreover, we take the opportunity to simplify the formalism introduced in~\cite{collapse}.
The present paper is intended as a self-contained introduction to collapse models stemming from the
theory of causal fermion systems.

It is an outstanding open problem of contemporary physics to reconcile the linear dynamics
of quantum theory with the reduction of the state vector in a measurement process.
Different solutions to this so-called {\em{measurement problem}} have been proposed,
among them Bohmian mechanics~\cite{duerr+teufel}, the many-worlds interpretation~\cite{dewitt-manyworlds},
decoherence~\cite{joos} and collapse models~\cite{gao}.
In {\em{collapse models}} one regards the wave function as the fundamental physical object of the theory,
whereas the particle character is explained from the dynamical collapse during a measurement process.
The reduction of the wave function is realized by modifying the Schr\"odinger dynamics with nonlinear
and stochastic terms. This idea was first made precise in the {\em{Ghirardi-Rimini-Weber (GRW)
model}}~\cite{ghirardi1, pearle0, ghirardi-pearle-rimini, ghirardi, ghirardi2}.
Other prominent examples of collapse models are the {\em{Di\'osi-Penrose (DP) model}}~\cite{diosi1, diosi2, penrose-collapse} and the {\em{continuous spontaneous localization (CSL) model}}~\cite{pearle0,
ghirardi-pearle-rimini}. For related works and surveys see also~\cite{bassi-ghirardi, pearle, bassi-duerr-hinrichs}.

In recent years, collapse models have gained increasing attention because they 
predict deviations from conventional quantum
theory to be tested in experiments~\cite{bassi, donadi-piscicchia, piscicchia2}.
These experiments are based on the fact that the stochastic term in the Schr\"odinger equation
gives rise to a heating of the probe, leading to emission of photons.
The experimental setup is to try to detect these photons; see~\cite{carlesso}
for a recent review.

In collapse models, the modifications of the Schr\"odinger dynamics are usually introduced
ad hoc. One idea to explain why the collapse comes about, being at the heart
of the DP model, is to consider gravity as being fundamentally classical, and to associate the nonlinear
correction term to the nonlinear coupling of the wave function in Newton's law of gravity
(see also~\cite{karolyhazy, karolyhazy2, penrose-collapse2}).
The theory of causal fermion systems yields a different explanation for why collapse occurs.
In this mechanism, collapse is again triggered by the coupling to bosonic fields.
But now, the field is not the gravitational field, but rather a multitude of fields which are specific
to the causal fermion system approach.
We model this multitude of fields in the non-relativistic limit as a stochastic background formed of
Gaussian and Markovian fields. Making use of the nonlinear nature of the interaction, one obtains
an effective dynamical collapse model~\cite{collapse}. This model has many similarities with the CSL model,
but also features significant differences. In simple terms, these differences can be understood from the fact that
the multitude of fields couple to the Schr\"odinger wave function in a nonlocal way both in space and time,
on a very small length scale~$\ell_{\min}$ which can be associated to the
Planck length (for details see Section~\ref{secdirnonloc}).
Consequently, conservation laws (like the conservation
of probability) are not formulated via spatial integrals, but rather in terms of integrals over spacetime strips,
referred to as {\em{surface layers}}. As we shall work out in detail, the energy of the probe can be expressed
in terms of a corresponding surface layer integral. Moreover, this {\em{energy is conserved in the statistical mean}}.
This means that, in contrast to the CSL model, in our model the stochastic background does not give rise
to an overall heating of the probe. 
Consequently, our collapse model does {\em{not}} predict the emission of photons in the collapse experiments.

In order to clarify the connection between the collapse models, we point out that, just as the CSL model,
our model gives rise to an evolution of the statistical operator of the standard Lindblad form,
except that the operators in the double-commutator are nonlocal in space and time
on the scale~$\ell_{\min}$.
But our stochastic evolution equation for the quantum state is quite different from the CSL model.
This is not a contradiction if one keeps in mind that the dynamical equation for the statistical operator
is obtained by taking a statistical mean. This procedure involves a loss of information, because
the stochastic degrees of freedom are integrated out. This makes the transition 
from the stochastic dynamical equation of the state to the Lindblad equation
non-unique, in the sense that different stochastic evolution equations for the quantum state can give rise
to the same Lindblad equation for the statistical operator. Our collapse model gives an explicit example
of this non-uniqueness: Although the stochastic evolution equation for the state is quite different
from the CSL model, including different physical predictions like the no-heating property,
both models give rise to the same Lindblad-type equation.

We now describe our methods and results in some more detail in words.
In the theory of causal fermion systems, the physical equations are formulated via a novel
variational principle in spacetime, the so-called {\em{causal action principle}}.
The resulting linearized field equations were analyzed in detail in~\cite{nonlocal}.
It was discovered that the linearized fields have a surprisingly rich structure.
In addition to homogeneous classical fields (like plane electromagnetic waves), there are many
additional fields, which all propagate with the speed of light. 
These fields do not all couple to all electrons in the same way, but instead the coupling depends
on the momentum of the electrons. In mathematical terms, all these fields are described by
a {\em{non-local integral operator}} in the Dirac equation (for details see~\eqref{dirnonloc} in Section~\ref{secdirnonloc}).
Another specific feature of the causal fermion system approach which is crucial for our collapse mechanism
is that all spacetime structures are encoded in the family of all physical wave functions.
This means in particular that, in contrast to for example Dirac wave functions in the presence of
external fields, the scalar product on the wave functions depends on the external fields.
It is formulated in terms of a surface layer integral (for details see~\eqref{c11}--\eqref{c3} in Section~\ref{dirnonloc}).
Describing the multitude of fields stochastically, one gets into the situation where not only all the
wave functions, but also the scalar product itself depend on the stochastic background.
The time evolution is such that the scalar product is preserved for each configuration of the background fields
(i.e.\ even before the stochastic mean is taken).
This setup is fundamentally different from the usual description of collapse models
with a stochastic Schr\"odinger equation, in which case the scalar product is fixed as the standard spatial
$L^2$-scalar product and only the wave functions are modified by the stochastic potential.
It is the main conclusion of the present paper that this difference has experimental consequences,
because our collapse mechanism does not necessarily lead to a heating of the probe.
Consequently, our model does {\em{not}} predict radiation caused by the
heating of the probe.

In order to obtain this result, we proceed in several steps.
In preparation, the formalism introduced in~\cite{collapse} is enhanced to the so-called equal-time formalism. The advantage of this approach is that it is closer to the standard formalism of stochastic PDEs and makes the combinatorics of the expansion more transparent.
In this enhanced formalism, it becomes possible to compute how the energy of the probe changes due to the
interaction with the environment. We show that no heating occurs in the statistical mean
(Theorem~\ref{thmheating}). Our findings are discussed and compared in detail with other collapse models.

The paper is organized as follows. After providing the necessary preliminaries (Section~\ref{secprelim}),
the equal time formalism is developed (Section~\ref{secequaltime}).
Heating is analyzed (Section~\ref{secheating}) and discussed in comparison 
with general heating results (Section~\ref{secnogo}), the CSL model
(Section~\ref{seccsl}), models involving quantized background fields
(Section~\ref{secquantum}) and the ETH formulation of quantum theory 
(Section~\ref{secETH}).

\section{Preliminaries} \label{secprelim}
The theory of {\em{causal fermion systems}} is a recent approach to fundamental physics (see the reviews~\cite{dice2014, review}, the textbooks~\cite{cfs, intro} or the website~\cite{cfsweblink}).
For brevity, here we shall not introduce the general setup (a brief general introduction streamlined to what is needed for collapse phenomena is given in~\cite{collapse}).
Instead, we merely explain a few general features of the approach
in words and introduce those objects which will be needed for our analysis.

\subsection{A Few General Features of Causal Fermion Systems }\label{secbasics}
We now explain a few features of the causal fermion system approach in words.
in this approach, spacetime and all structures therein are described
mathematically by a single object: a measure~$\rho$ on a set of linear operators
on a Hilbert space~$(\H, \la .|. \ra_\H)$.
The Hilbert space can be thought of as the usual one-particle Hilbert space of
quantum mechanics. By ``measure'' we mean a measure in the
sense of integration theory; i.e., a mapping which to
a subset~$V$ of the linear operators assigns a positive number~$\rho(V)$, which can be thought of as the ``volume'' of the set~$V$.
One advantage of working with such abstract measures is that
the measure~$\rho$ can take two roles. First, it distinguishes a
set of linear operators where it is non-zero. This statement is made mathematically precise
by the notion of the support of the measure, denoted by
\[ M:= \supp \rho \:. \]
The points of~$M$ are associated with the spacetime points, and~$M$ is
our {\em{spacetime}}. Thus in the causal fermion system approach, the spacetime
points are operators. These operators encode information on the local correlations
of all one-particle wave functions in our Hilbert space.
The second role of the measure~$\rho$ is that it provides an integration measure
on spacetime, which corresponds to the usual four-dimensional volume
measure~$d\rho|_M = dt\, d^3x$.

The physical equations are formulated in a causal fermion system via
a nonlinear variational principle for the measure~$\rho$, referred to as
the {\em{causal action principle}}. The corresponding {\em{Euler-Lagrange (EL)
equations}} tell us about the dynamics of the system.
In order to put this dynamics into the context of collapse theories, we need to
say a few general words on the general structure of these EL equations.
First, they are {\em{nonlinear}}. This means in particular
that there is no superposition principle for macroscopic physical systems.
There are limiting cases where one gets the linear dynamics of quantum
theory (see for example the recent papers~\cite{fockdynamics, qftlimit}).
But this linear dynamics holds only approximately; it is overruled by nonlinear
effects, similar as is the case in standard collapse theories.
In order to make this point clearer, we note that the dynamics of single wave
functions is described by the {\em{dynamical wave equation}}
\[ Q \psi = \mathfrak{r} \psi \:, \]
where~$Q$ is an integral operator in spacetime
(for details see~\cite{dirac, nonlocal}).
This equation resembles the Dirac equation and generalizes this equation to
our setting. Similar to the Schr\"odinger equation, it is a linear equation for the
wave function~$\psi$. However, it is nonlinear because the operator~$Q$ depends
on all the other wave functions of the system as well.

Next, we need to explain what {\em{causality}} means in our setting.
On the fundamental level, a causal fermion system provides the structures which allows us to distinguish whether
two spacetime points are spacelike or timelike separated,
and in the latter case which one lies in the future of the other.
(see~\cite[Sections~1.2 and~1.3]{cfs}).
But this does not mean that strong causation holds. On the contrary,
in general the future may affect the past on microscopic scales.
Causation holds only on macroscopic scales.
These statements can be made precise using positivity properties of
boundary integrals of lens-shaped spacetime regions (for details see~\cite{linhyp, dirac}).
We do not need to enter the details. But what is important in view of our collapse model is
that the dynamics is {\em{deterministic on scales}} which are {\em{much larger than the
Planck length}}, in the sense that initial data given at an initial time give rise to a unique solution of the EL equations at a later time.
But clearly, this determinism holds only if the
initial data is known. In the presence of a stochastic background (as will be considered
here), we only have a probability distribution for the initial data, implying that
also the time evolution becomes probabilistic.

Finally, in the context of collapse models, it is important to specify in which sense
we have {\em{finite speed of propagation}}. As just mentioned, on macroscopic
scales (i.e.\ scales much larger than the Planck length)
the Cauchy problem has a unique solution. Choosing Cauchy surfaces 
as the boundaries of lens-shaped regions, it also follows that the solutions
propagate with finite speed (details on this method can be found
in~\cite[Section~6.1]{dirac}). However, this statement is true only if the
solutions are evaluated weakly on the boundary (in terms of so-called
surface layer integrals).
Using a more physical language, finite speed of propagation is not seen
directly from the solutions of the equations, but becomes apparent
only if a position measurement
is performed. Such a position measurement is implemented not simply by evaluating
the wave function at a single spacetime point, but it corresponds to a suitable
integral expression involving the wave function. This feature of causal fermion
systems is also present in our effective collapse model and plays an important
role for understanding what no-faster-than-light signaling means in our model.
We will come back to this point in Section~\ref{secnogo}.

\subsection{The Nonlocal Dirac Equation} \label{secdirnonloc}
As pointed out in the previous section, a spacetime points of a causal
fermion system is a linear operator on a Hilbert space which encodes
information on the local correlations of all one-particle wave functions of the system.
These one-particle wave functions are also referred to as the {\em{physical wave functions}}. With this notion, one can say that in the causal fermion system approach, 
all spacetime structures are encoded in the family of all physical wave functions.
In the vacuum in Minkowski space, the physical wave functions are simply
formed of all negative-energy solutions of the Dirac equation
\beq \label{Dirvac}
(i \Pdd-m)\, \psi = 0 \:.
\eeq
In the interacting situation, on the other hand,
the dynamics of the family of physical wave functions is determined by the causal action principle.
As was worked out in detail in~\cite{nonlocal}, this dynamics differs from the standard Dirac dynamics.
It can be described effectively by inserting a {\em{nonlocal potential}}~$\B$ into the Dirac equation.
More precisely, the Dirac equation becomes
\beq \label{dirnonloc}
\big( i \Pdd + \B - m \big) \psi = 0 \:,
\eeq
where~$\B$ is an integral operator in spacetime with integral kernel~$\B(x,y)$, i.e.\
\[ 
\big( \B \psi \big)(x) = \int_M \B(x,y)\, \psi(y)\: d^4y \:. \]
The integral kernel~$\B(x,y)$ is formed of the bosonic potentials.
Their structure and dynamics is determined by the causal action principle as well.
An important finding of~\cite{nonlocal} is that~$\B$ is not only formed of the usual classical
potentials (like the electromagnetic potential). Instead, it involves a multitude of
potentials which all propagate with the speed of light.
Their coupling to the Dirac spinors depends on the momenta of the spinors.
Mathematically, this is described by the structure of the integral operator
\beq \label{Bstochastic}
\B(x,y) = \sum_{a=1}^N B_a \Big( \frac{x+y}{2} \Big) \:L_a(y-x)\:,
\eeq
where~$B_a$ are multiplication operators acting on the spinors and the~$L_a$ are 
complex-valued functions.
The potential~$\B$ is nonlocal both in space and time on a scale~$\ell_{\min}$
which lies between the Planck scale~$\varepsilon$ and the
length scale~$\ell_\text{macro}$ of macroscopic physics,
\beq \label{ellmindef}
\varepsilon \ll \ell_{\min} \ll \ell_\text{macro} \:.
\eeq
By ``nonlocal on the scale~$\ell_{\min}$'' we mean that the kernels~$L_a(y-x)$
vanish unless all the components~$|(y-x)^j|$ are smaller than~$\ell_{\min}$.
The number~$N$ of these potentials is very large and scales like
\[ 
N \simeq \frac{\ell_{\min}}{\varepsilon} \:. \]
The nonlocal potential is always {\em{symmetric}} with respect to the usual pointwise
inner product on Dirac spinors in Minkowski space. More precisely, denoting this inner product by
\[ 
\Sl \psi | \phi \Sr := \overline{\psi} \phi = \psi^\dagger \gamma^0 \phi \:, \]
(where the dagger denotes transposition and complex conjugation; $\overline{\psi}$ is sometimes referred to
as the adjoint spinor), the symmetry of the nonlocal
potential means that
\beq \label{Bsymm}
\Sl \psi \,|\, \B(x,y)\, \phi \Sr = \Sl \B(y,x)\, \psi \,|\, \phi \Sr \qquad \text{for all~$\psi, \phi \in \C^4$}\:.
\eeq

As already mentioned, in a causal fermion system, all spacetime structures
are encoded in the family of physical wave functions. This also means that the familiar
structures are not give a-priori, but must be constructed from this family of wave functions.
In particular, this is the case for the scalar product on the physical wave functions,
which is preserved in time and gives the usual connection to the probabilistic interpretation of
the quantum mechanical wave functions. It turns out that this scalar product
is no longer simply a spatial integral
but has a more complicated structure, depending on the whole family of physical wave functions
(the abstract description can be found in~\cite{dirac} or~\cite[Section~9.4]{intro}).
In our effective description by the nonlocal Dirac equation~\eqref{dirnonloc}, this scalar
product, denoted by~$\la .|. \ra_t$, takes the form
\begin{align}
\la \psi | \phi \ra_t &:= \int \Sl \psi \,|\, \gamma^0\, \phi \Sr \: d^3x \label{c11} \\
&\quad\;\; -i \int_{x^0<t} d^4x \int_{y^0>t} d^4y\;
\Sl \psi(x) \,|\, \B(x,y)\, \phi(y) \Sr \label{c2} \\
&\quad\;\; +i \int_{x^0>t} d^4x \int_{y^0<t} d^4y\;
\Sl \psi(x) \,|\, \B(x,y)\, \phi(y) \Sr \:. \label{c3}
\end{align}
Note that~\eqref{c11} is the usual scalar product on Dirac wave functions.
The additional summands~\eqref{c2} and~\eqref{c3} can be understood as correction terms
which take into account the nonlocality of the potential~$\B$ in~\eqref{dirnonloc}.
The mathematical structure of these additional terms is that of a so-called {\em{surface layer integral}}.
A direct computation using~\eqref{Bsymm} yields that the inner product~$\la \psi | \phi \ra_t$
is indeed time independent (for details see~\cite[Proposition~B.1]{baryogenesis}).

We close with a few clarifying remarks:
\bitem
\item[{\rm{(a)}}] It is a specific feature of the
causal fermion system approach that the form of the {\em{scalar product depends
on the bosonic potentials}}.
This feature will be crucial for our collapse mechanism, as will be
explained in Section~\ref{secnogo}.
\item[{\rm{(b)}}]
We point out that the dynamics described by the nonlocal Dirac 
equation~\eqref{dirnonloc} is linear. On the other hand, as elaborated
in~\cite{gisin, gisin2, gisin-rigo}, it is crucial for collapse models
that there is also a {\em{nonlinearity}} which triggers the collapse.
The same is true in our model: A nonlinearity is needed in order to overrule
the superposition principle, giving rise to a reduction of the state vector.
The way to think of this nonlinearity in our setting is
that the potential~$\B$ depends on all the physical wave functions of the system,
making the equation~\eqref{dirnonloc} implicitly nonlinear.
At present, it is not clear how to bring this nonlinearity into an explicit 
and more tractable form; this is a project of future research.
A more detailed
discussion of this point is given in~\cite[Section~4.9]{collapse}.
\item[{\rm{(c)}}] In the present paper, we only consider the system on the
level of one-particle quantum mechanics. As explained in~\cite[Section~4.7]{collapse},
our results also carry over to {\em{many-body systems}} described in Fock spaces.
Moreover, the amplification effect in the number of particles applies as usual.
Apart from this general effect, a precise estimate of the collapse time has not
yet been worked out. This would also make it necessary to quantify the
nonlinear effects mentioned under~(b), going beyond the scope of the
present paper.
\item[{\rm{(d)}}] In the present paper, we restrict attention to the non-relativistic
limit. The above setting also gives rise to a corresponding
{\em{relativistic collapse model}}.
The details of this model are subject of current research.
For the time being, we refer the reader interested in relativistic extensions
of our model to the outlook in~\cite[Section~5]{collapse}.
\eitem

\subsection{Hamiltonian Formulation}
Our main task is to analyze how the nonlocal potential affects the dynamics of the Dirac waves.
To this end, it is most convenient to write the Dirac equation~\eqref{dirnonloc} in the Hamiltonian form.
We let~$\H_t = L^2(\R^3, \C^4)$ be the Hilbert space of square integrable Dirac wave functions at time~$t$. 
We denote the scalar product on this Hilbert space by
\beq \label{L2sp}
( \psi | \phi )_t := \int \Sl \psi \,|\, \gamma^0\, \phi \Sr \big|_{(t,\vec{x})} \: d^3x \:.
\eeq
We introduce the operator~$V(t,t') : \H_{t'} \rightarrow \H_{t}$ by
\[ 
(V(t,t') \psi)(\vec{x}) = \int_{\R^3} \Big(- \gamma^0 \, \B\big( (t,\vec{x}), (t', \vec{y}) \Big) \: \psi(\vec{y})\: d^3y \:. \]
This operator is symmetric in the sense that
\beq \label{Vsymm}
V(t,t')^\dagger = V(t',t) \:. 
\eeq
Now the Dirac equation can be written in the Hamiltonian form as
\[ 
i \partial_t \psi = H \psi \:, \]
where the Hamiltonian is of the form
\beq \label{H0V}
H = H_0 + V 
\eeq
with the standard Dirac Hamiltonian~$H_0 := -i \gamma^0 \vec{\gamma} \vec{\nabla} + m \gamma^0$
and the interaction potential
\[ (V \psi)(t) = \int_{-\infty}^\infty V(t,t')\: \psi(t')\: dt' \:. \]

Following the procedure in~\cite{collapse}, one could remove the free Hamiltonian~$H_0$
from the equation of motion by transforming to the interaction picture.
Here we prefer not to work in the interaction picture, because for the analysis of heating
in Section~\ref{secheating} it is preferable to keep~$H_0$ in the dynamical equations.

The Dirac equation~\eqref{H0V} is nonlocal in time. This is a major deviation from the standard formulation
of physics. As a consequence, it is no longer obvious how to propagate a solution to the future, because,
intuitively speaking, on the time scale~$\ell_{\min}$ the ``future can affect the past.''
Nevertheless, as is made precise in the recent paper~\cite{cauchynonloc},
the time evolution operator exists and is unique, provided that~$\ell_{\min}$ 
and the amplitudes of the waves~$B_a$ are sufficiently small.
More computationally, the time evolution operator can be expressed in terms of a nonlocal Dyson series
(for details see~\cite[Section~4.2]{collapse}).
Here we do not need to enter the details of this analysis. It suffices to write the 
resulting time evolution operator as
\beq \label{Utt0}
U^t_{t_0} \::\: \H_{t_0} \rightarrow \H_t \:, \qquad \psi(t_0) \mapsto \psi(t) \:.
\eeq
We point out that this time evolution operator is {\em{not}} unitary; i.e.\ it does in general
{\em{not}} preserve the Hilbert space scalar product~\eqref{L2sp}.
Instead, only the modified scalar product~$\la .|. \ra_t$ involving the surface layer integrals
in~\eqref{c11}--\eqref{c3} is preserved in time.

\section{The Equal Time Formalism} \label{secequaltime}
In~\cite{collapse} the dynamics of the Dirac waves was analyzed using the nonlocal Dyson series.
Here we shall proceed a bit differently in the so-called {\em{equal time formalism}}.
This formalism is mathematically equivalent, but has the advantage that it is a bit simpler
and closer to the standard formulation of collapse models.

\subsection{The Equal Time Hamiltonian} $\quad$
Employing the time evolution operator in~\eqref{H0V}, we can rewrite the nonlocal operator~$V(t,t')$
as an operator acting on the wave function at time~$t$,
\beq \label{Wdef}
\big( V \psi \big)(t) = W(t)\, \psi_t \qquad \text{with} \qquad W(t)  = \int_{-\infty}^\infty V(t,t')\: U^{t'}_t\: dt' \:.
\eeq
Working with the operator~$W$, the Dirac equation in the Hamiltonian form takes the familiar form
\beq \label{schW}
i \partial_t \psi_t = \big( H_0 + W(t) \big)\, \psi_t \:.
\eeq
It can be solved with the standard Dyson series. More precisely, the time evolution operator in~\eqref{Utt0}
can be written as the ordered exponential
\beq \label{Texp}
U^t_{t_0} = \Pexp \bigg( -i \int_{t_0}^t e^{-i (t-\tau) H_0} \:W(\tau)\: e^{-i (\tau-t) H_0} \: d\tau \bigg) \: e^{-i (t-t_0) H_0} \:,
\eeq
which can then easily be expanded in a power series in~$W$.
We refer to the above formulation as the {\em{equal time formalism}}.
It has the advantage that we can work with the standard Dyson series~\eqref{Texp}.
Moreover, by taking the Gaussian pairings in the right way
(as will be explained in detail in Section~\ref{secgauss}), the computations can be greatly simplified.
We refer to the operator~$W$ in~\eqref{Wnosymm} as the {\em{equal time Hamiltonian}}
(in the interaction picture). Note that the equal time Hamiltonian is in general
{\em{not}} symmetric with respect to the standard $L^2$-scalar product, because
\beq \label{Wnosymm}
W(t)^\dagger = \int_{-\infty}^\infty (U^{t'}_t)^\dagger \,V(t',t) \overset{\text{in general}}{\neq} W(t)\:.
\eeq

In the equal time formalism, we can express the conserved scalar product~\eqref{c11}--\eqref{c3}
in terms of the standard $L ^2$-scalar product by
\beq \label{Sigdef}
\la \psi | \phi \ra_t = \big( \psi \,\big|\, (\1+\Sig_t)\, \phi \big)_t \qquad \text{for all~$\psi, \phi \in \H_m$}\:,
\eeq
where the operator~$\Sig_t$ is defined by
\beq \label{Sigt}
\Sig_t := -\frac{i}{2} \int_{-\infty}^\infty d\tau \int_{-\infty}^\infty d\tau' \;
\big( \epsilon(\tau-t) - \epsilon(\tau'-t) \big)\: (U^\tau_t)^\dagger \: V^\tau_{\tau'} \: U^{\tau'}_t
\eeq
(where~$\epsilon$ is the sign function defined by~$\epsilon(x)=1$ if~$x>0$ and~$\epsilon(x)=-1$ otherwise).
The operator~$\Sig_t$ is obviously symmetric with respect to the standard $L^2$-scalar product,
\[ 
\Sig_t^\dagger = \Sig_t \:. \]

In order to illustrate the obtained formalism, we now verify by a detailed computation that the 
scalar product, written as on the right side of~\eqref{Sigdef}, is preserved in time. 
\begin{Lemma} In the equal time formalism,
\begin{gather}
\partial_t \Sig_t =  i (\1+\Sig_t) (H_0+W) - i (H_0+W^\dagger) (\1+\Sig_t) \label{dtSig} \\
\partial_t \big( \psi \,\big|\, (\1+\Sig_t) \phi \big)_t = 0 \:. \label{dtsprod}
\end{gather}
\end{Lemma}
\Proof
In order to derive~\eqref{dtSig}, we first differentiate the time evolution operator in~\eqref{Texp}
with respect to~$t$ and~$t_0$, one finds that
\[ \partial_t U^t_\tau = -i (H_0+W)\: U^t_\tau \qquad \text{and} \qquad \partial_t U^\tau_t = i  U^\tau_t\: (H_0+W) \:. \]
These formulas make it possible to compute the time derivative of~$\Sig_t$ in~\eqref{Sigt} by
\begin{align*}
\partial_t \Sig_t &= -\frac{i}{2} \int_{-\infty}^\infty d\tau \int_{-\infty}^\infty d\tau' \;
\big( -2 \delta(\tau-t) +2 \delta (\tau'-t) \big)\: (U^\tau_t)^\dagger \: V(\tau.\tau') \: U^{\tau'}_t \\
&\quad\: -\frac{i}{2} \int_{-\infty}^\infty d\tau \int_{-\infty}^\infty d\tau' \;
\big( \epsilon(\tau-t) - \epsilon(\tau'-t) \big)\: \partial_t \Big( (U^\tau_t)^\dagger \: V(\tau, \tau') \: U^{\tau'}_t \Big) \\
&= i \int_{-\infty}^\infty d\tau' \;V(t,\tau') \: U^{\tau'}_t 
-i \int_{-\infty}^\infty d\tau\: (U^\tau_t)^\dagger \: V(\tau, t) \\
&\quad\: -\frac{i}{2} \int_{-\infty}^\infty d\tau \int_{-\infty}^\infty d\tau' \;
\big( \epsilon(\tau-t) - \epsilon(\tau'-t) \big)\\
&\qquad\quad \times \Big( -i (H_0+W^\dagger) \:(U^\tau_t)^\dagger \: V(\tau, \tau') \: U^{\tau'}_t
+ i (U^\tau_t)^\dagger \: V(\tau, \tau') \: U^{\tau'}_t\: (H_0+W) \Big) \:.
\end{align*}
This formula can be simplified by using~\eqref{Wdef} and~\eqref{Sigt}. We thus obtain
\begin{align*}
\partial_t \Sig_t &=  i (H_0+W) - i (H_0+W^\dagger) - i (H_0+W^\dagger)\, \Sig_t + i \Sig_t (H_0+W)  \:, 
\end{align*}
proving~\eqref{dtSig}.

Next, we differentiate~\eqref{Sigdef}, apply the chain rule and
use the Schr\"odinger equation~\eqref{schW}. This gives
\begin{align*}
\partial_t \big( \psi \,\big|\, (\1+\Sig_t) \phi \big)_t
&= i \big( (H_0+W) \psi \,\big|\, (\1+\Sig_t) \phi \big)_t \\
&\quad\; - i \big( \psi \,\big|\, (\1+\Sig_t) (H_0+W) \phi \big)_t
+ \big( \psi \,\big|\, \dot{\Sig}_t \phi \big)_t = 0 \:,
\end{align*}
concluding the proof.
\QED

Finally, similar to~\cite[Section~4.3]{collapse}, we transform to the standard $L^2$-scalar product by setting
\beq \label{tildepsi}
\tilde{\psi}(t) =\sqrt{\1+\Sig_t}\: \psi(t) \:.
\eeq
Then the Schr\"odinger equation~\eqref{schW} can be written as
\beq
i \partial_t \tilde{\psi}(t) = \big( H_0 + \tilde{W} \big) \tilde{\psi}(t) \label{schWtilde}
\eeq
with the new equal time Hamiltonian
\[ 
H_0 + \tilde{W} := (\1+\Sig_t)^\frac{1}{2} \,(H_0+W)\, (\1+\Sig_t)^{-\frac{1}{2}}
- i (\1+\Sig_t)^\frac{1}{2} \Big( \partial_t (\1+\Sig_t)^{-\frac{1}{2}} \Big) \:. \]
At this stage, the operator~$\tilde{W}$ looks somewhat complicated, but
its form can be simplified considerably (see Lemma~\ref{lemmaWtil} below). Here we merely note that, by construction, we know that
\begin{align*}
0 &= \partial_t \big( \psi \,\big|\, (\1+\Sig_t) \phi \big)_t
= \partial_t ( \tilde{\psi} \,|\, \tilde{\phi} )_t
= i \:\big( (H_0+\tilde{W}) \tilde{\psi} \,\big|\, \tilde{\phi} \big)_t - i \:\big(\tilde{\psi} \,\big|\,  \big( H_0 + \tilde{W}
\big) \tilde{\phi} \big)_t \:,
\end{align*}
showing that, in contrast to the potential~$W$ in~\eqref{Wnosymm}, the transformed potential~$\tilde{W}$ is indeed symmetric with respect to the standard $L^2$-scalar product,
\beq \label{Wtsymm}
\tilde{W}^\dagger = \tilde{W}\:.
\eeq

\subsection{Expansion in Powers of $\ell_{\min}\,\|\B\|$}
The general formulas derived in the previous section simplifies considerably if we use the fact that the
nonlocal potential is of short time range. We denote this time range by~$\ell_{\min}$.
In the equal time formulation, each time integral appearing in the definition of the equal time operators
(like the integrals in~\eqref{Wdef} and~\eqref{Sigt}) gives a scaling factor~$\ell_{\min}$.
With this in mind, we can perform an expansion in powers of the dimensionless parameter
\[ \ell_{\min}\,\|\B\| \:, \]
which we assume to be much smaller than one.
In order to capture collapse phenomena, we need to expand to second order.

\begin{Lemma} \label{lemmaWtil}
The operator~$\tilde{W}$ in the transformed Schr\"odinger equation~\eqref{schWtilde}
has the expansion
\beq \label{Wtil}
\tilde{W} = \frac{1}{2}\, (W + W^\dagger)
-\frac{1}{8}\: \big[ (W-W^\dagger),  \Sig_t \big] + \O \big( \ell_{\min}^2\, \|\B\|^3 \big) \:.
\eeq
\end{Lemma} \noindent
We note that the right side of this equation is obviously symmetric, in agreement with our
general observation~\eqref{Wtsymm}. We also note that the detailed form of the error term
can be understood from the following scaling argument.
The potential~$\tilde{W}$ (and similarly~$\B$, $W$ and~$W^\dagger$) have the
scaling dimension~$1/\text{length}$
(as is obvious from~\eqref{schWtilde}). Moreover, $\ell_{\min}$ is the only length dimension
which enters our expansion. This determines the power of either~$\|\B\|$ or~$\ell_{\min}$ in~\eqref{Wtil}.
Having this scaling argument in mind, in the following computations it suffices to expand in powers
of~$\ell_{\min}$ (of course, with a little bit of extra work, one could also keep track of the
corresponding powers of the potential).
\Proof[Proof of Lemma~\ref{lemmaWtil}]
According to~\eqref{Sigt}, the operator~$\Sig_t$ is of order~$\O( \ell_{\min}\,\|\B\|)$. Therefore,
\begin{align}
\tilde{\psi} &= \sqrt{1+\Sig_t}\, \psi = \Big( \1 + \frac{\Sig_t}{2}\Big) \psi + \O \Big( \big( \ell_{\min}\, \|\B\| \big)^2 \Big) 
\psi
\label{tilpsipsi} \\
\psi &= (1+\Sig_t)^{-\frac{1}{2}}\, \tilde{\psi} = \Big( \1 - \frac{\Sig_t}{2}\Big) \tilde{\psi}
+ \O \Big( \big( \ell_{\min}\, \|\B\| \big)^2 \Big) \psi
\:. \label{psitilpsi}
\end{align}
When taking the time derivative of~\eqref{tilpsipsi}, one must be careful because, according to~\eqref{dtSig},
differentiating~$\Sig_t$ decreases the order in~$\ell_{\min}$ by one, because
(again using that~$\Sig_t=\O( \ell_{\min}\,\|\B\|)$)
\[ \partial_t \Sig_t = i (W - W^\dagger) + \O\big( \ell_{\min}\,\|\B\|^2 \big)\:. \]
Therefore, we need to expand the square root one order higher,
\begin{align*}
\partial_t \tilde{\psi} &= \partial_t \bigg( \Big( \1 + \frac{\Sig_t}{2} -\frac{1}{8}\: \Sig_t^2 \Big) \psi \bigg)
+ \O \big( \ell_{\min}^2\, \|\B\| ^3 \big) \psi \\
&= \Big( \1 + \frac{\Sig_t}{2} \Big) \partial_t \psi + \frac{\partial_t \Sig_t}{2}\: \psi
-\frac{1}{8} \Big( (\partial_t \Sig_t) \Sig_t + \Sig_t\: (\partial_t \Sig_t) \Big) \Big) \psi
+ \O \big( \ell_{\min}^2\, \|\B\| ^3 \big) \psi \:.
\end{align*}
Now we can employ~\eqref{schW} as well as~\eqref{dtSig} and expand in powers of the
potential. We thus obtain
\begin{align*}
&\partial_t \tilde{\psi} 
= -i \Big( \1 + \frac{\Sig_t}{2}\Big) (H_0+W) \psi \\
&\qquad\: + \frac{i}{2} \:\Big( (\1+\Sig_t) (H_0+W) - (H_0+W^\dagger) (\1+\Sig_t) \Big)\, \psi  + \O \big( \ell_{\min}^2\, \|\B\| ^3 \big) \psi \\
&= -i H_0 \psi -\frac{i}{2}\, (W + W^\dagger) \psi 
-\frac{i}{2}\:\Sig_t\, (H_0+W) \psi + \frac{i}{2} \:\Sig_t\, (H_0+W) \,\psi
- \frac{i}{2}\: (H_0+W^\dagger) \,\Sig_t\, \psi \\
&\quad\:-i (W-W^\dagger)\, \frac{\Sig_t}{8}\:\psi -i \,\frac{\Sig_t}{8}\,(W-W^\dagger)\,\psi + \O \big( \ell_{\min}^2\, \|\B\|^3 \big) \psi \\
&= -i H_0 \psi - \frac{i}{2}\: H_0\, \Sig_t\, \psi -\frac{i}{2}\, (W + W^\dagger) \psi 
- \frac{i}{2}\:W^\dagger \,\Sig_t\, \psi \\
&\quad\: -i (W-W^\dagger)\, \frac{\Sig_t}{8}\:\psi -i \,\frac{\Sig_t}{8}\,(W-W^\dagger)\,\psi + \O \big( \ell_{\min}^2\, \|\B\|^3 \big) \psi \\
&= -i H_0 \psi - \frac{i}{2}\: H_0\, \Sig_t\, \psi -\frac{i}{2}\, (W + W^\dagger) \psi \\
&\quad\: - \frac{i}{8}\: W \Sig_t \psi - \frac{i}{8}\: \Sig_t W \psi 
- \frac{3i}{8}\: W^\dagger \Sig_t\, \psi +\frac{i}{8}\: \Sig_t W^\dagger \psi
+ \O \big( \ell_{\min}^2\, \|\B\|^3 \big) \psi\:.
\end{align*}
Next, we use~\eqref{psitilpsi} to re-express~$\psi$ in terms of~$\tilde{\psi}$,
\begin{align*}
&\partial_t \tilde{\psi} = -iH_0 \tilde{\psi} -\frac{i}{2}\, (W + W^\dagger) \tilde{\psi} + \frac{i}{4}\, (W + W^\dagger) \,\Sig_t\, \tilde{\psi}\\
&\quad\:- \frac{i}{8}\: W \Sig_t \tilde{\psi} - \frac{i}{8}\: \Sig_t W \tilde{\psi} 
- \frac{3i}{8}\: W^\dagger \Sig_t\, \tilde{\psi} +\frac{i}{8}\: \Sig_t W^\dagger \tilde{\psi} + \O \big( \ell_{\min}^2\, \|\B\|^3 \big) \psi \\
&= -\frac{i}{2}\, (W + W^\dagger) \tilde{\psi} 
+ \frac{i}{8}\: W \Sig_t \tilde{\psi} - \frac{i}{8}\: \Sig_t W \tilde{\psi} 
- \frac{i}{8}\: W^\dagger \Sig_t\, \tilde{\psi} +\frac{i}{8}\: \Sig_t W^\dagger \tilde{\psi} + \O \big( \ell_{\min}^2\, \|\B\|^3 \big) \psi \\
&= -\frac{i}{2}\, (W + W^\dagger)\, \tilde{\psi} 
+ \frac{i}{8}\: \big[ (W-W^\dagger),  \Sig_t \big] \,\tilde{\psi} + \O \big( \ell_{\min}^2\, \|\B\|^3 \big) \psi \:.
\end{align*}
Comparing with~\eqref{schWtilde} gives the result.
\QED

\subsection{Taking the Statistical Mean} \label{secgauss}
According to~\eqref{Bstochastic}, the potential~$\B$, and therefore also the
equal time Hamiltonians~$\tilde{W}$ and~$W$, involve stochastic potentials~$B_1, \ldots B_L$.
When computing the density operator, we need to take the statistical mean over the stochastic potentials.
Denoting the density operator by~$\sigma_t$ and the statistical mean by~$\bbra \cdots \kket$,
we have
\beq \label{rhotildedef}
\sigma_t := \bbra \:\big| \tilde{\psi}(t) \big) \big( \tilde{\psi}(t) \big| \:\kket \:,
\eeq
where~$\tilde{\psi}$ is the transformed wave function~\eqref{tildepsi}. 
As in~\cite[Section~4.4]{collapse}, we assume that these stochastic potentials are Gaussian and Markovian, i.e.\
\begin{align}
\bbra \gamma^0 B_a (t, \vec{x}) \kket &= 0 \label{markov1} \\
\bbra \:\big(\gamma^0 B_a(t, \vec{x}) \big)^\alpha_\beta\: 
\big( \gamma^0 B_b(t',\vec{y}) \big)^\gamma_\delta \:\kket &= \delta(t-t')\: \delta_{ab}\:
C^{\;\:\,\alpha \gamma}_{a, \beta \delta}(\vec{x}, \vec{y}) \:. \label{markov2}
\end{align}
Here for simplicity we restrict attention to a covariance which does not depend on time
(the case with time dependence can be treated similarly, as was worked out to some extent in see~\cite{collapse}).
This makes it possible to derive dynamical equations for the density operator.
At this stage we do not need to enter the details, but it suffices to recall the general procedure.
The dynamics of~$\tilde{\psi}$ is expressed in the equal time formalism by the Schr\"odinger equation~\eqref{schWtilde}, which can be solved similar to~\eqref{Texp} with an ordered exponential,
\beq \label{Utildett0}
\begin{split}
\tilde{\psi}(t) &= \tilde{U}^t_{t_0}\: \tilde{\psi}(t_0) \qquad \text{with} \\
\tilde{U}^t_{t_0} \,&\!:= \Pexp \bigg( -i \int_{t_0}^t e^{-i (t-\tau) H_0} \:\tilde{W}(\tau)\: e^{-i (\tau-t) H_0} \: d\tau \bigg)\: e^{-i (t-t_0) H_0} \:.
\end{split}
\eeq
Finally, the equal time Hamiltonian~$\tilde{W}$ is given more explicitly in Lemma~\ref{lemmaWtil}.
Using these formulas in~\eqref{rhotildedef}, we obtain the time evolution of the projection operator~$| \tilde{\psi}(t) ) ( \tilde{\psi}(t)|$. Taking the statistical mean and using~\eqref{markov1} and~\eqref{markov2},
one gets sums of terms involving Gaussian pairings
(basics on Gaussian fields, the covariance, the Wick rule and resulting Gaussian pairings
can be found for example in~\cite[Section~6.2]{glimm+jaffe}).

The combinatorics of the Gaussian parings simplifies considerably when expanding
in powers of~$\ell_{\min} \|\B\|$. Namely, to leading order in~$\ell_{\min}\, \|\B\|$,
it suffices to take into account those contributions where there is at most one Gaussian pairing
within every time strip of width~$\sim \ell_{\min}$.
In the equal time formalism, this implies that in every operator of second order in the potential,
the Gaussian pairings must combine these two potentials. Moreover, third and higher powers of the
potentials can be omitted. For example, the commutator in~\eqref{Wtil} can be simplified
with the following replacement rule,
\[ -\frac{1}{8}\: \big[ (W-W^\dagger),  \Sig_t \big] \;\mapsto\;
-\frac{1}{8}\: \bbra \big[ (W-W^\dagger),  \Sig_t \big] \kket
+ \O \big( \ell_{\min}^2\, \|\B\|^3 \big) \:. \]
The right side is a symmetric potential which is no longer stochastic. Therefore, it is irrelevant for
collapse phenomena. Instead, it can be regarded as a correction to the free Dirac dynamics.
Combining this potential with the free Hamiltonian~$H_0$, it no longer appears in the interaction picture.
In this way, we conclude that the second summand in~\eqref{Wtil} may be disregarded.
Consequently, using~\eqref{Wdef}
and~\eqref{Vsymm}, we may replace~$\tilde{W}$ by its linear expansion
\beq \label{Wtsimp}
\tilde{W}(t) = \int_{-\infty}^\infty \tilde{W}^{t'}(t)\: dt' + \O \big( V^2 \big)
\eeq
with
\beq \label{Wtdef}
\tilde{W}^{t'}(t) :=\frac{1}{2}\: \big( V(t,t')\: e^{-i (t'-t)H_0} + e^{-i (t-t')H_0}\: V(t',t) \big) \:.
\eeq

Before working out the dynamics with Gaussian pairings, it is convenient to diagonalize the covariance.
Here we merely give the result of this construction, referring for details to~\cite[Lemma~3.1
and Section~4.4]{collapse}. In an eigenbasis of the covariance,
we have new Gaussian random fields~$W_{k,a}(t)$ which are independent and normalized,
\[ 
\bbra W_{k,a}(t) \kket = 0 \qquad \text{and} \qquad
\bbra W_{k,a}(t)\, W_{l,b}(t') \kket = \delta(t-t')\, \delta_{ab} \, \delta_{kl} \:, \]
where~$k$ is a new quantum number labelling the eigenvalues of the covariance.
Moreover, the symmetrized potential in~\eqref{Wtdef} can be written as
\beq \label{Vrep2}
\tilde{W}^{t'}(t) = \sum_{k,a} \: M_{k,a}(t-t')\: W_{k,a}\Big( \frac{t+t'}{2} \Big)
\eeq
where~$M_{k,a}(t-t')$ are (for any fixed~$t$ and~$t'$) spatial operators having the symmetry properties
\[ 
M_{k,a}(t-t') = M_{k,a}(t'-t) = M_{k,a}(t-t')^\dagger \:. \]
Note that the first relation follows immediately from the fact that on the left side of~\eqref{Vrep2}
we symmetrized in~$t$ and~$t'$, whereas the second relation is a consequence of the
symmetry property of the nonlocal potential~\eqref{Bsymm}.

\subsection{The Lindblad Dynamics}
We now want to derive the resulting dynamics of the density operator.
\begin{Prp} The density operator satisfies the dynamical equation
\beq \label{lindblad}
\begin{split}
\frac{d}{dt}\: \sigma_t &= -i \big[ H_0, \sigma_t \big] -\sum_{k,a} 
\int_{-\infty}^\infty d\zeta  \int_0^\infty d\nu \:
\Big[ M_{k,a}(\zeta), \big[M_{k,a} \big( \zeta - \nu \big), \sigma_t \big] \Big] \\
&\quad\: + \O \big( \ell_{\min}^2\, \|\B\|^3 \big)\:\sigma_t \:.
\end{split}
\eeq
\end{Prp}
\Proof We return to the ordered exponential~\eqref{Utildett0}.
Differentiating with respect to~$t$, we obtain
\[ \partial_t \tilde{\psi}(t) = -i \big(H_0+\tilde{W}(t) \big)\: \tilde{U}^t_{t_0}\, \tilde{\psi}(t_0) \:. \]
Taking the statistical mean, the random fields in~$\tilde{W}(t)$ must be paired with corresponding
fields in the Dyson series~$\tilde{U}^t_{t_0}$. To the considered leading order in~$\ell_{\min}\,\|\B\|$,
it suffices to take into account the pairing with the potential which appears in the Dyson series at the very left,
being evaluated at the time of the outermost integral. Combining all the other integrals again to a
Dyson series,
we obtain
\begin{align*}
\bbra \partial_t \tilde{\psi}(t) \kket &= \big( -i H_0 + A(t) \big)\: \bbra \tilde{\psi}(t) \kket \qquad \text{with} \notag \\
A(t) \,&\!:= \bbra \big(-i \tilde{W}(t)\big) \bigg( -i \int_{-\infty}^t \: \tilde{W}(\tau)\: d\tau \bigg) \kket\:,
\end{align*}
with error terms of the order~$\O \big( \ell_{\min}^2\, \|\B\|^3 \big)$.
Using the formulas~\eqref{Wtsimp}--\eqref{Vrep2}, taking the expectation value with Gaussian pairings
gives (again up to errors of order~$\O(\ell_{\min}^2\, \|\B\|^3)$)
\begin{align}
A(t) &=-\bbra \int_{-\infty}^\infty dt'\: \tilde{W}^{t'}(t)
\bigg( \int_{-\infty}^t d\tau \int_{-\infty}^\infty d\tau'\: \tilde{W}^{\tau'}(\tau) \bigg) \kket \label{Aform} \\
&= -\sum_{k,a} 
\int_{-\infty}^\infty dt' M_{k,a}(t-t') 
\int_{-\infty}^t d\tau \int_{-\infty}^\infty d\tau'\: M_{k,a}(\tau-\tau') \: \delta\Big( \frac{t+t'}{2}-\frac{\tau+\tau'}{2} \Big) \notag \\
&= -2 \sum_{k,a} 
\int_{-\infty}^\infty dt'  \int_{-\infty}^t d\tau \:M_{k,a}(t-t')\: M_{k,a} \big( 2\tau-t-t' \big)
= \left\{ \begin{array}{lc} 
\zeta := t-t' \notag \\
\nu := 2(t-\tau)
\end{array} \right\} \notag \\
&= - \sum_{k,a} 
\int_{-\infty}^\infty d\zeta  \int_0^\infty d\nu \:M_{k,a}(\zeta)\: M_{k,a} \big( \zeta - \nu \big) \:. \label{Acomp}
\end{align}

We now move on to the density operator~$\sigma_t$ defined by~\eqref{rhotildedef}.
Taking the time derivative, one can take the Gaussian pairings as explained above. The only difference
is that one must also take into account pairings between bra and ket.
A direct computation gives the result.
\QED

The dynamical equation~\eqref{lindblad} resembles the {\em{Lindblad equation}}
(sometimes referred to more accurately as the Gorini-Kossakowski-Sudarshan-Lindblad equation),
which is usually written in the form
\beq \label{standardlindblad}
\frac{d}{dt}\: \sigma_t = -i \big[ H_0, \sigma_t \big] -\sum_{\kappa} 
\Big( L_\kappa^\dagger L_\kappa\, \sigma_t - 2 L_\kappa \sigma_t L_\kappa^\dagger
+  \sigma_t\, L_\kappa^\dagger L_\kappa \Big) \:.
\eeq
One difference is that, in our setting, the operators~$M_{k,a}$ in the double
commutator are nonlocal in space and time on the scale~$\ell_{\min}$.
Another major difference is that, in contrast to~\eqref{standardlindblad},
the operators~$M_{k,a}(\zeta)$ and~$\big[M_{k,a} \big( \zeta - \nu \big)$ in~\eqref{lindblad}
do not need to be adjoints of each other. The equation~\eqref{lindblad} also preserves the trace
of the statistical operator, but in general it does not preserve positivity.
Nevertheless, as is explained in detail in~\cite[Section~4.4]{collapse},
under sensible general technical assumptions it is possible to transform and reparametrize
the operators in~\eqref{lindblad} such as to obtain a
Lindblad-type equation of the form (see~\cite[Theorem~4.1]{collapse})
\[ \frac{d \sigma_t}{dt} = -i [H, \sigma_t] - \frac{1}{2} \sint_\varkappa 
\big[ K_\varkappa, [K_\varkappa, \sigma_t] \big] \; \Big( 1 + \O \big( \ell_{\min}\, \|\B\| \big) \Big)\:. \]

\subsection{Energy Conservation} \label{secheating}
We now come to the question of whether the interaction with the
stochastic background field gives rise to a heating of the probe.
By ``heating'' we mean the effect that the energy of the probe increases
due to the interaction with the environment. This energy is given by the
statistical mean of the expectation value of the free Hamiltonian,
\beq \label{Efree}
E = \bbra ( \psi | H_0 \psi ) \kket \:.
\eeq
Here the difficulty arises that the stochastic background field modifies the form
of the scalar product. In order to get into the position where the energy change
can be computed in a clean way, as in~\cite[Section~4.6]{collapse}
we consider an idealized situation where the stochastic background field vanishes
in a time strip near the initial time~$t_0$ and the final time~$t_1$, as is shown in Figure~\ref{figscatter}.
\begin{figure}
\psset{xunit=.5pt,yunit=.5pt,runit=.5pt}
\begin{pspicture}(404.98344842,244.77890856)
{
\newrgbcolor{curcolor}{0.80000001 0.80000001 0.80000001}
\pscustom[linestyle=none,fillstyle=solid,fillcolor=curcolor]
{
\newpath
\moveto(1.00606564,171.6087706)
\lineto(403.98184894,172.0567958)
\lineto(403.98188674,61.71779738)
\lineto(1.49995162,61.3536021)
\closepath
}
}
{
\newrgbcolor{curcolor}{0.80000001 0.80000001 0.80000001}
\pscustom[linewidth=2.0031495,linecolor=curcolor]
{
\newpath
\moveto(1.00606564,171.6087706)
\lineto(403.98184894,172.0567958)
\lineto(403.98188674,61.71779738)
\lineto(1.49995162,61.3536021)
\closepath
}
}
{
\newrgbcolor{curcolor}{0 0 0}
\pscustom[linewidth=2.75905519,linecolor=curcolor]
{
\newpath
\moveto(25.24584945,0.0134421)
\lineto(23.03101606,227.39768525)
}
}
{
\newrgbcolor{curcolor}{0 0 0}
\pscustom[linestyle=none,fillstyle=solid,fillcolor=curcolor]
{
\newpath
\moveto(15.34365056,223.45994606)
\lineto(22.86171448,244.77890843)
\lineto(30.79362672,223.61043636)
\curveto(26.25619859,227.4291001)(19.82514601,227.36645851)(15.34365056,223.45994606)
\closepath
}
}
{
\newrgbcolor{curcolor}{0 0 0}
\pscustom[linewidth=2.0031495,linecolor=curcolor]
{
\newpath
\moveto(1.25281512,19.98260242)
\lineto(403.73064567,20.80736336)
}
}
{
\newrgbcolor{curcolor}{0 0 0}
\pscustom[linewidth=2.0031495,linecolor=curcolor]
{
\newpath
\moveto(1.25281512,212.60324619)
\lineto(403.73060787,213.42800714)
}
\rput[bl](-23,10){$t_0$}
\rput[bl](-20,205){$t_1$}
\rput[bl](50,30){$\Sig_t=0$}
\rput[bl](50,180){$\Sig_t=0$}
\rput[bl](50,110){$\Sig_t \neq 0$}
}
\end{pspicture}
\caption{Energy conservation. The shaded region indicates where the
collapse-inducing stochastic field is active.}
\label{figscatter}
\end{figure}%
Thus, similar to a scattering process, we want to understand how the energy
changes from time~$t_0$ to time~$t_1$. Since the stochastic background field vanishes
at initial and final times, we may replace~\eqref{Efree} by the expectation
value in the presence of~$\B$,
which can be expressed alternatively in the untransformed and the transformed variables,
\beq \label{Et}
\begin{split}
E(t) \;&\!:= \bbra\; \big\la \psi(t) \,\big|\, \big(H_0+W(t) \big) \, \psi(t) \big\ra_t \;\kket \\
&= \bbra \big( \tilde{\psi}(t) \,\big|\, \big(H_0+\tilde{W}(t) \big) \, \tilde{\psi}(t) \big) \kket \:.
\end{split}
\eeq
The physical meaning of~$E(t)$ at intermediate times is not quite clear, and we do not
need to discuss it here. For us, the formula~\eqref{Et} gives us an interpolation of the
energy at initial and final times, defined such that the time evolution can be
analyzed.

\begin{Thm} \label{thmheating} The stochastic Schr\"odinger equation~\eqref{schWtilde} does not
give rise to heating, in the sense that the expectation value of the energy does not change in the
statistical mean up to a small error,
\beq \label{Econs}
\frac{d}{dt}\,E(t) = \O \big( \ell_{\min}^2\, \|\B\|^3 \big) \:E(t)\:.
\eeq
\end{Thm}
\Proof Differentiating~\eqref{Et}
by~$t$ and using~\eqref{schWtilde} as well as the symmetry property~\eqref{Wtsymm},
we obtain
\[ \dot{E}(t) = \bbra \big( \tilde{\psi}(t) \,\big|\, \big( \partial_t \tilde{W}(t)\big) \, \tilde{\psi}(t) \big) \kket \:. \]
The stochastic potentials contained in~$\partial_t \tilde{W}(t)$ must be paired
with potentials either in the bra or the ket vector. A computation similar to~\eqref{Acomp} gives
\beq \label{Edot}
\dot{E}(t) = \tr \big( (B(t) + B(t)^\dagger)\, \sigma^t \big)
\eeq
with (again up to errors of order~$\O(\ell_{\min}^2\, \|\B\|^3)$)
\begin{align*}
B(t) &=-i \:\bbra \int_{-\infty}^\infty dt'\: \big( \partial_t \tilde{W}^{t'}(t) \big)\,
\bigg( \int_{-\infty}^t d\tau \int_{-\infty}^\infty d\tau'\: \tilde{W}^{\tau'}(\tau) \bigg) \kket \\
&=-i \:\frac{d}{dt} \:\bbra \int_{-\infty}^\infty dt'\: \tilde{W}^{t'}(t)\:
\bigg( \int_{-\infty}^t d\tau \int_{-\infty}^\infty d\tau'\: \tilde{W}^{\tau'}(\tau) \bigg) \kket \\
&\quad\: + i \:\bbra \int_{-\infty}^\infty dt'\: \tilde{W}^{t'}(t)\:
\frac{\partial}{\partial t} \bigg( \int_{-\infty}^t d\tau \int_{-\infty}^\infty d\tau'\: \tilde{W}^{\tau'}(\tau) \bigg) \kket \:.
\end{align*}
Comparing with~\eqref{Aform}, we obtain
\beq \label{Bres}
B(t) = i \dot{A}(t) + i \:\bbra \bigg( \int_{-\infty}^\infty dt'\: \tilde{W}^{t'}(t)\:
\int_{-\infty}^\infty d\tau'\: \tilde{W}^{\tau'}(t) \bigg) \kket \:.
\eeq
According to~\eqref{Acomp}, the operator~$A(t)$ is time independent.
Moreover, in view of the prefactor~$i$, the last summand in~\eqref{Bres} is obviously anti-symmetric.
We conclude that~$B(t) + B(t)^\dagger$ vanishes, and using this result in~\eqref{Edot}
concludes the proof.
\QED

We finally discuss the size of the error term in~\eqref{Econs}. 
Clearly, this error term is not equal to zero. But it is by a
factor~$\ell_{\min}^2 \|\B\|^2$ smaller than the energy transfer
in standard collapse models. In other words, in our collapse model
heating is suppressed at the order where collapse occurs.
In this way, our mechanism reduces the heating of the probe by many orders of magnitude
far beyond experimental bounds. This justifies why we can talk of
``no heating.''

\subsection{Wave Function Collapse}
We now explain our collapse mechanism. We again consider the idealized
situation shown in Figure~\ref{figscatter} where the stochastic background field vanishes
in a time strip near the initial time~$t_0$ and the final time~$t_1$.
Then the collapse can be detected by analyzing how the variance of an observable~$\mathcal{O}$ changes in time. More
precisely, our task is to analyze the expression
\[ \big( \bbra \la \mathcal{O}^2 \ra \kket - \bbra \la \mathcal{O} \ra^2 \kket \big) \big|_{t_0}^{t_1} \:. \]
Here~$\la . \ra$ denotes the expectation value.
Due to our assumption that~$\B$ vanishes in the initial and final time strips, this expectation value
can be computed equivalently with the untransformed or transformed scalar products.
Considering the corresponding scalar product for intermediate times and using the fundamental theorem
of calculus, we have the choice to analyze alternatively
\[ \text{either} \quad \frac{d}{dt}  \Big( \bbra ( \tilde{\psi} |  \mathcal{O}^2 \tilde{\psi} ) \kket - \bbra ( \tilde{\psi} | 
\mathcal{O} \tilde{\psi} )^2 \kket \Big) \quad \text{or} \quad 
\frac{d}{dt}  \Big( \bbra \la \psi |  \mathcal{O}^2 \psi \ra_t \kket - \bbra \la \psi |  \mathcal{O} \psi \ra_t^2 \kket \Big)  \:. \]
It is instructive to analyze both alternative descriptions and to discuss their relations.

In either description, we need to discuss two contributions: The one from the expectation value of~$\mathcal{O}^2$
and the one from the square of the expectation value of~$\mathcal{O}$.
We begin with the first contribution: The expectation value of~$\mathcal{O}^2$ involves fluctuations of the
stochastic fields. These fluctuations drop out when taking the statistical mean, which is typically small.
Moreover, the statistical mean typically does not have a preferred sign
(in contrast to the statistical mean of the energy, as will be discussed in Section~\ref{seccsl} below).
For this reason, following the standard procedure in CSL models (see~\cite{gisin, bassi-ghirardi})
we set this expectation value to zero for every~$t$.
Then it remains to consider the contribution by the square of the expectation value of~$\mathcal{O}$; i.e.,
\beq \label{square}
\text{either} \qquad -\frac{d}{dt}  \bbra ( \tilde{\psi} |  \mathcal{O} \tilde{\psi} )^2 \kket \qquad \text{or} \qquad 
- \frac{d}{dt}  \bbra \la \psi |  \mathcal{O} \psi \ra_t^2 \kket \:.
\eeq
Here it makes a major difference which scalar product is used. 
In the first case, we can apply~\eqref{schWtilde}
\[ \frac{d}{dt}  ( \tilde{\psi} |  \mathcal{O} \tilde{\psi} ) =\big( -i (H_0 + \tilde{W}) \tilde{\psi} \,\big|\,  \mathcal{O} \tilde{\psi} \big)
+ \big( \tilde{\psi} \,\big|\,  \mathcal{O} (-i (H_0+\tilde{W}) \tilde{\psi} \big) \]
and use the symmetry of the potential~$\tilde{W}$ in~\eqref{Wtsymm} to obtain the expectation value of a
commutator,
\[ 
\frac{d}{dt}  ( \tilde{\psi} |  \mathcal{O} \tilde{\psi} ) = i ( \tilde{\psi} |  \big[ (H_0 + \tilde{W}), \mathcal{O}\big] \tilde{\psi} ) \:. \]
Taking the derivative of the square in~\eqref{square} gives
\beq \label{c1a}
- \frac{d}{dt}  ( \tilde{\psi} |  \mathcal{O} \tilde{\psi} )^2 = -2 \:
( \tilde{\psi} |  \mathcal{O} \tilde{\psi} )\: \frac{d}{dt}  ( \tilde{\psi} |  \mathcal{O} \tilde{\psi} ) 
= -2 i\: ( \tilde{\psi} |  \mathcal{O} \tilde{\psi} )\: ( \tilde{\psi} |  \big[ (H_0 + \tilde{W}), \mathcal{O}\big] \tilde{\psi} ) \:.
\eeq
Taking the statistical mean, the potential~$\tilde{W}$ in the last expectation value
can be paired with the potential in the first expectation value.
This gives a contribution
\beq \label{c12}
-\frac{d}{dt}  \bbra ( \tilde{\psi} |  \mathcal{O} \tilde{\psi} )^2 \kket \simeq
- \bbra \big| ( \tilde{\psi} |  \big[ \tilde{W}, \mathcal{O}\big] \tilde{\psi} ) \big|^2 \kket \:.
\eeq
Being negative, this contribution describes a reduction of the wave function, provided that the
commutator~$[ (H_0 + \tilde{W}), \mathcal{O}]$ has a non-zero expectation value.

In the second case, on the other hand, we can apply~\eqref{schW},
\begin{align}
\frac{d}{dt}  \la \psi |  \mathcal{O} \psi \ra_t &= \la -i (H_0 + W) \psi |  \mathcal{O} \psi \ra_t
+ \la \psi |  \mathcal{O} (-i (H_0+W) \psi \ra_t \notag \\
&= \frac{i}{2}\: \la \psi |  \big[ (2 H_0 + W + W^\dagger) , \mathcal{O}\big] \psi \ra_t
-\frac{i}{2} \: \la \psi |  \big\{ (W - W^\dagger), \mathcal{O} \big\} \psi \ra_t \label{case2}
\end{align}
(where the curly brackets denote the anti-commutator).
Now the expectation value of the derivative of the square can be computed similar to~\eqref{c1a} and~\eqref{c12}.
We get a contribution involving the square of the terms in~\eqref{case2}. More precisely,
\beq \label{c22}
-\frac{d}{dt}  \bbra \la \psi |  \mathcal{O} \psi \ra_t^2 \kket \simeq
- \frac{1}{4}\: \bbra \Big| \la \psi \,|\, \big[ (W + W^\dagger) , \mathcal{O}\big]\, \psi \ra_t - 
\la \psi \,|\, \{ (W + W^\dagger) , \mathcal{O}\big\}
\, \psi \ra \Big|^2 \kket \:.
\eeq

At first sight, the results of these computations are quite different: On the one hand, in~\eqref{c12}, 
the non-vanishing of the commutator is crucial for the collapse.
In~\eqref{c22}, on the other hand, one can get collapse even if the commutator term vanishes.
In this case, the collapse is triggered by the fact that the potential~$W$ is not symmetric.
These two findings are compatible if one keeps in mind that the transformation~\eqref{tildepsi}
from~$\psi$ to~$\tilde{\psi}$ involves the operator~$\Sig_t$, which is itself stochastic.
Moreover, the square root~$\sqrt{\1 + \Sig_t}$ is a nonlocal spatial operator.
In~\cite[Section~4.6]{collapse} it is argued using causal propagation and locality of the measurement
apparatus that it is sensible to assume that the commutator in the untransformed equation~\eqref{c22}
vanishes (this corresponds to the usual assumptions in the CSL model; see for example~\cite{bassi-ghirardi}).
Under this assumption, the collapse occurs as a consequence of the fact that~$W$ is not symmetric.
Alternatively, working with~\eqref{c12}, one can argue that, as a consequence of the stochasticity
and non-locality of the transformed potential~$\tilde{W}$, the commutator in~\eqref{c12} is non-zero,
again giving rise to collapse.

\section{Connection to General Theorems Proving Spontaneous Heating} \label{secnogo}
In the recent paper~\cite{donadi-ferialdi-bassi} it was shown
under general assumptions that
collapse models necessarily give rise to a diffusion and thus a heating of the probe. We now explain why
these general results do not apply to our model. The results
in~\cite{donadi-ferialdi-bassi} are based on the following assumptions:
\bitem
\item[{\rm{(i)}}] There is no superluminal signaling.
\item[{\rm{(ii)}}] The model is space-translation covariant, at least at the
statistical level.
\eitem
The assumption of no-faster-than-light signalling poses strong constraints
on the structure of the collapse model. Indeed, combined with the assumptions
of Markovian fields and of a linear dynamics of the density operator,
it gives rise to a dynamical equation for the density operator of Lindblad
form~\cite{bassi-duerr-hinrichs}.

At least at first sight, our model seems to comply with the above assumptions.
In particular, the Dirac equation~\eqref{Dirvac}
is hyperbolic and yields finite propagation
speed. Moreover, it is sensible to assume that the covariance in~\eqref{markov2}
is translation invariant (i.e., that it depends only on the difference vector~$\vec{x}-\vec{y}$).
Therefore, our finding of no heating seems to contradict the results
in~\cite{donadi-ferialdi-bassi}. We now explain why there is no contradiction.

Before beginning, we point out that
one needs to make a subtle distinction between ``diffusion'' and ``heating''.
For the non-relativistic Hamiltonian $H_0=p^2/2m$, diffusion implies that the momentum
changes and thus the expectation value of~$p^2$ increases, also giving rise to an
increase of the expectation value of the energy. In our model, however, the operator~$H_0$
is the Dirac Hamiltonian, which is unbounded below. Therefore, the connection between
diffusion and an energy increase is not so clear. In particular, diffusion does not necessarily
give rise to heating. While our results rule out heating, they do not currently
make a statement about diffusion.

In order to explain why the results of~\cite{donadi-ferialdi-bassi} do not immediately
apply to our setting, we need to explain in detail what we precisely mean
by ``finite propagation speed'' and ``superluminal signaling.''
First of all, one must keep in mind that the nonlocal potential~$\B$
in the Dirac equation~\eqref{dirnonloc} in general
destroys finite propagation speed. Indeed, the results on finite propagation speed
(as can be obtained for example by using energy estimates for symmetric
hyperbolic systems; see for example~\cite[Chapter~5.3]{john}
or~\cite[Chapter~13]{intro}) break down once a nonlocal operator is included.
On the other hand, as mentioned in Section~\ref{secbasics}, in causal fermion
systems there is finite propagation speed on macroscopic scales
(as worked out in~\cite[Section~4]{linhyp} and~\cite[Section~6.1]{dirac}).
But this finite speed of propagation is not seen by looking at the wave functions
pointwise, but only by evaluating an integral expression which can be understood
physically as describing a position measurement.
In our effective collapse model, such a position measurement
must be performed by computing an expectation value with respect to the scalar
product~$\la .|. \ra_t$ which involves surface layer integrals
(see~\eqref{c11}--\eqref{c3}). These surface layer integrals have the effect that
possible contributions to the wave functions which violate causal propagation
drop out and are no longer visible in the scalar product~$\la .|. \ra_t$.
However, working instead with the standard probability integral
(i.e., taking only the spatial integral in~\eqref{c11} and disregarding the
surface layer integrals~\eqref{c2} and~\eqref{c3}), these causality-violating
contributions are visible, thereby violating causal propagation.

This point can also be understood from a somewhat different perspective
by transforming one scalar product to the other. Indeed,
using the transformation~$\psi \mapsto \tilde{\psi} = \sqrt{\1 + \Sig_t} \psi $ 
in~\eqref{tildepsi}, we could arrange that the scalar product takes the standard form.
But this transformation is nonlocal, and therefore causal propagation is no longer
visible when working with~$\tilde{\psi}$. Therefore,
in order to analyze whether superluminal 
signalling occurs, one must necessarily work with the untilde wave functions and
the scalar product~$\la .|. \ra_t$.

To summarize, our model exhibits no superluminal signaling, as becomes apparent
by considering expectation values of position operators with respect to the
physical scalar product~$\la .|. \ra_t$. But, in contrast to the standard setting
of collapse theories, this scalar product is nonlocal and involves the stochastic
background field. With this in mind, the no-faster-than-light signalling is stated
in a way different from the assumption~(i) above. This is why the results
of~\cite{donadi-ferialdi-bassi} do not apply to our collapse model.

\section{Comparison with the CSL Model} \label{seccsl}
It is generally believed that the Lindblad equation gives rise to a heating of the probe
(see for example~\cite{piscicchia2}). We now recall the standard argument in the simplest possible
setting and discuss it afterward. We assume that our initial state~$\psi$ is the ground state of the free Hamiltonian~$H_0$, i.e.\
\[ H_0 \psi = E \psi \qquad \text{with} \qquad E = \inf \sigma(H_0) \:. \]
Moreover, we consider the dynamics as described by the Lindblad equation~\eqref{standardlindblad}.
Then the statistical operator~$\sigma_t = |\psi \ra \la \psi|$ satisfies at initial time the equation
\begin{align}
&\frac{d}{dt}\: \bbra \la \psi | H_0 \psi \ra \kket = \frac{d}{dt}\: \bbra \la \psi | (H_0-E) \psi \ra \kket \label{rel1} \\
&= \Tr_\H \bigg\{ (H_0-E) \bigg( -i \big[ H_0, \sigma_t \big] -\sum_{\kappa} 
\Big( L_\kappa^\dagger L_\kappa\, \sigma_t - 2 L_\kappa \sigma_t L_\kappa^\dagger
+  \sigma_t\, L_\kappa^\dagger L_\kappa \Big) \bigg) \bigg\} \\
&= -\sum_{\kappa} \Big\la \psi \,\Big|\, \Big(  (H_0-E)\,  L_\kappa^\dagger L_\kappa
- 2 L_\kappa^\dagger (H_0-E)\,  L_\kappa
+ L_\kappa^\dagger L_\kappa\, (H_0-E)  \Big) \psi \Big\ra \\
&= 2 \sum_{\kappa} \la \psi \,|\, L_\kappa^\dagger (H_0-E)\,  L_\kappa \psi \ra
= 2 \sum_{\kappa} \la L_\kappa \psi \,|\, (H_0-E)\,  L_\kappa \psi \ra \geq 0 \:. \label{rel4}
\end{align}
Here in~\eqref{rel1} we used that the Lindblad dynamics preserves the trace of the statistical operator.
In~\eqref{rel4} we used the eigenvalue equation~$(H_0-E)\psi=0$ as well as the fact that the
operator~$H_0-E$ is positive.

Let us discuss the range of validity of this argument. We first point out that
the computation~\eqref{rel1}--\eqref{rel4} holds in large generality. It is mathematically sound
(functional analytic subtleties could be avoided by restricting attention to a sufficiently large
but finite-dimensional subspace of the Hilbert space). One objection is that it holds only in linear approximation,
but does not apply in a fully nonlinear description. On the other hand, the considered linear approximation seems
justified if the stochastic background field is sufficiently weak. With this in mind, the perturbative
description seems sufficient, and we will stick to it in what follows.
Then the only shortcoming which needs to be discussed is the assumption that~$\psi$ is a
{\em{ground state}} of~$H_0$. This assumption is justified for a bound state of the Schr\"odinger equation.
However, the Dirac Hamiltonian is not bounded from below, so that it is a-priori not clear what a bound
state should be.

In order to make sense of the concept of a ground state, in the quantum mechanical
description one projects out the negative-energy states of~$H_0$. After doing so, the Hamiltonian is
bounded from below, so that the notion of a ground state becomes sensible.
The physical picture behind this procedure is that the negative-energy states do not take part in the
interaction and can therefore be disregarded. In the quantum-field-theoretic description, the negative-energy
states are treated by working with a Hartree-Fock ground state where all the negative-energy states are occupied
(as envisioned by Dirac in his Dirac sea picture), and by describing the system relative to this ground state
by acting with creation operators for particle and anti-particles.

The reason why the assumption of a ground state is not sensible in our collapse model
is that the negative-energy states do take in part in the interaction. Indeed, the nonlocal potential
in the Dirac equation~\eqref{dirnonloc} couples to all solutions of the Dirac equation, including the solutions
negative energy. Therefore, the negative-energy states must not be disregarded; on the contrary,
they play an important role for the dynamics of an electron.
Intuitively speaking, the energy transferred to an electron by
the mechanism described in~\eqref{rel4} is compensated by an energy transfer from the electron
to the states of the Dirac sea. This is why the argument based on the computation~\eqref{rel1}--\eqref{rel4} is
too naive and does not apply in our setting. Instead, the energy change must be computed
as in Theorem~\ref{thmheating}. The conclusion is that, in the collapse model derived in the
setting of causal fermion systems, the energy of the probe is conserved in the statistical mean.
Since physical measurements involve taking a statistical mean, this means that in experiments
no heating of the probe occurs, and consequently that the probe should not emit radiation due to
a heating by the stochastic background.

A general lesson of the above consideration is that a Lindblad-type equation by itself does not determine
whether heating occurs or not. In addition, one needs to specify the dynamical equation
for the state as well as the precise form of the conservation laws.

\section{Comparison with Models Involving Quantized Background Fields} \label{secquantum}
Our model also has similarities with collapse models involving
quantized background field (see~\cite{pearle-completely, bedingham, pearle-stress, fu}
and applications in~\cite{piccione-bassi1, donadi-bassi, nobakht, piccione-bassi2}).
We now discuss the similarities and differences to our collapse model.
The similarities become most apparent if one considers the limiting case 
worked out in~\cite{fockdynamics} where
the nonlocal potential~$\B$ in the Dirac equation~\eqref{dirnonloc} goes over
to field operators of a second-quantized bosonic field (see also the review~\cite{qftlimit}).
In this limiting case, our collapse model goes over to a model of fermions
in the presence of a second-quantized bosonic field, just as considered in the
above-mentioned papers. In particular, the heating effect can be seen by
computing expectation values of the fermionic Hamiltonian~$H_0$.
The differences between our model and the models involving second-quantized
background fields become apparent if one goes beyond the limiting
case of bosonic fields. Indeed, without taking the limiting case considered
in~\cite{fockdynamics}, the causal action principle gives rise to nonlinear corrections
to the linear Fock dynamics. Moreover, the scalar product of the
one-particle Hilbert space~$\la .|. \ra_t$ in~\eqref{c11}--\eqref{c3} depends on the
stochastic background and involves nonlocal surface layer integrals.
As already explained in Section~\ref{secnogo}, these features play an essential role
in our collapse model.

\section{Comparison with the ETH Formulation of Quantum Theory} \label{secETH}
The ETH formulation of quantum theory is an approach to overcome the measurement problem
(see~\cite{blanchard2016garden, froehlich2019review, froehlich2019relativistic}. One key ingredient is the {\em{principle of diminishing potentialities}}, implemented mathematically by the observation
that the algebra generated by the future observables becomes smaller
as time evolves. This has the effect that at certain times, {\em{events}} form, being defined by the condition
that the center of the centralizer of the future algebra be non-trivial. At each event, the system splits into
two or more subsystems. This branching process leads to a {\em{tree}} of events.
The resulting density operator is shown to satisfy a Lindblad equation.
The ETH formulation has been studied in concrete models~\cite{froehlich-gang-pizzo, froehlich-pizzo,
froehlich-pizzo2}. We remark that there is a connection between the ``events'' of the ETH formulation
and the spacetime point operators of a causal fermion system; for details see~\cite{eth-cfs}.

So far, energy conservation has not been studied in connection with the ETH formulation.
It is also not obvious how to do so. The main difficulty is that, in the ETH formulation, the dynamics
of the state is recovered by unraveling the Lindblad dynamics.
However, as discussed in the previous section, this ``unraveling'' is not unique, because the
same Lindblad equation can be obtained from different stochastic equations for the state vector.
As one sees in the above examples of the stochastic Schr\"odinger equation of the CSL model
versus the dynamical equation coming from causal fermion systems, the choice of the stochastic equation
may well have an influence on whether energy conservation holds or not. Therefore, before one could study
heating phenomena in the ETH formulation, one would have to specify the dynamical equation for the state
vector when events occur (with ``events'' as defined in the ETH formulation via a non-triviality of
the center of the centralizer of the future algebra).

\Thanks{{{\em{Acknowledgments:}}
We are grateful to Catalina Curceanu, Lajos Di\'osi, J\"urg Fr\"ohlich, Kristian Piscicchia and Alessandro Pizzo
for helpful discussions. We would like to thank the participants of the conference ``Causal Fermion Systems 2025''
for valuable feedback. We are also grateful to the referees for their careful
reading and valuable comments.

\bibliographystyle{amsplain}
\providecommand{\bysame}{\leavevmode\hbox to3em{\hrulefill}\thinspace}
\providecommand{\MR}{\relax\ifhmode\unskip\space\fi MR }
\providecommand{\MRhref}[2]{%
  \href{http://www.ams.org/mathscinet-getitem?mr=#1}{#2}
}
\providecommand{\href}[2]{#2}

\end{document}